\documentclass[aps,prb,twocolumn,superscriptaddress,showpacs,floatfix]{revtex4}
\usepackage{graphicx}
\usepackage{latexsym}
\usepackage{stmaryrd}
\usepackage{amsmath}
\usepackage{amssymb}
\usepackage{epstopdf}

\begin{document}
\title{Engineering Topological quantum dot through planar magnetization in bismuthene}
\author{Jiaojiao Zhou}
\affiliation{School of Physical Science and Technology, Soochow University, Suzhou, 215006, China}
\author{Tong Zhou}
\affiliation{Department of Physics, University at Buffalo, State University of New York, Buffalo, New York 14260, USA}
\author{Shu-guang cheng}
\affiliation{Department of Physics, Northwest University, Xi'an 710069, People's Republic of China}
\author{Hua Jiang}\email{jianghuaphy@suda.edu.cn}
\affiliation{School of Physical Science and Technology, Soochow University, Suzhou, 215006, China}
\affiliation{Institute for Advanced Study, Soochow University, Suzhou 215006, China}
\author{Zhongqin Yang}
\affiliation{Department of Physics, State Key Laboratory of Surface Physics and Key Laboratory for Computational Physical Sciences (MOE), Fudan University, Shanghai 200433, China}
\date{\today}

\begin{abstract}
The discovery of quantum spin Hall materials with huge bulk gaps in experiment, such as bismuthene, provides a versatile platform for topological devices. We propose a topological quantum dot (QD) device in bismuthene ribbon in which two planar magnetization areas separate the sample into a QD and two leads. At zero temperature, peaks of differential conductance emerge, demonstrating the discrete energy levels from the confined topological edge states. The key parameters of the QD, the tunneling coupling strength with the leads and the discrete energy levels, can be controlled by the planar magnetization and the sample size. Specially, different from the conventional QD, we find that the angle between two planar magnetization orientations provides an effective way to manipulate the discrete energy levels. Combining the numerical calculation and the theoretical analysis, we identify that such manipulation originates from the unique quantum confinement effect of the topological edge states. Based on such mechanism, we find the spin transport properties of QDs can also be controlled.
\end{abstract}


\maketitle

\section{Introduction}
The Quantum spin Hall (QSH) effect with topological helical edge states is one of the focuses in the topological state studies \cite{CLKane,CLKane1,BABernevig,MKonig,SQShen,XLQi1,XLQi,MZHasan}. In the QSH systems, carriers can propagate without dissipation along the edge channels. At one edge,  carriers of spin-up and spin-down flow to opposite directions. Specifically, the propagation direction of the carriers of the same spin is reversed at the opposite edge. Such unique transport properties enable QSH systems to own promising applications in low-power electronics and spintronics\cite{YZhangKHe, YFRen, VSverdlovS, PChen}. In the early years, the studies of QSH systems mainly focus on group-VI monolayers (e.g., graphene\cite{CLKane,CLKane1} and silicene\cite{CCLiu,CCLiu2}) and semiconductor quantum wells (e.g., HgTe/CdTe\cite{BABernevig,MKonig} and InAs/GaSb\cite{CXLiu,IKnez}). However, the weak spin-orbit coupling of these systems only leads to a tiny bulk gap with the order of a few meV, which limits the applications in the topological devices. Recently, the QSH effect is proposed in group-V monolayers (e.g., bismuthene \cite{ZFWang,CCLiu3,IKDrozdov,XLi} and antimonene \cite{ZGSong,YDMa,SGCheng,MPumera}). In these systems, the bulk gap  can reach the same order of the atomic spin-orbit coupling strength of Bi and Sb. Significantly, F. Reis et al. have synthesized the bismuthene sample on the SiC substrate, and they observed a large topological bulk gap up to $0.8$ eV\cite{FReis}. The discovery generates extensive attentions in the QSH phase on group-V monolayers and the related applications in topological devices \cite{GLiWHanke,JGouBYXia,HQHuangFLiu}. Furthermore, it is found that  the topological  characteristics of group-V monolayers can easily be modified by doping, adsorption, chemical modification or substrate effect \cite{WXJi,SCChen,SHKim,AJMannix,FYang,LChen,BTFu,HGao,THirahara, THirahara1}. For example, the first-principles calculations have demonstrated that the magnetic doping or substrate can induce very large exchange fields (up to 400 meV) in the functionalized bismuthene\cite{TZhouJYZhang} and antimonene \cite{TZhou,TZhou1}, which drives the system from the QSH phase to quantum anomalous Hall and valley polarized QSH phases, respectively. This easily modified feature of group-V monolayer is favorable for building topological devices.

\begin{figure*}
\includegraphics[width=14 cm]{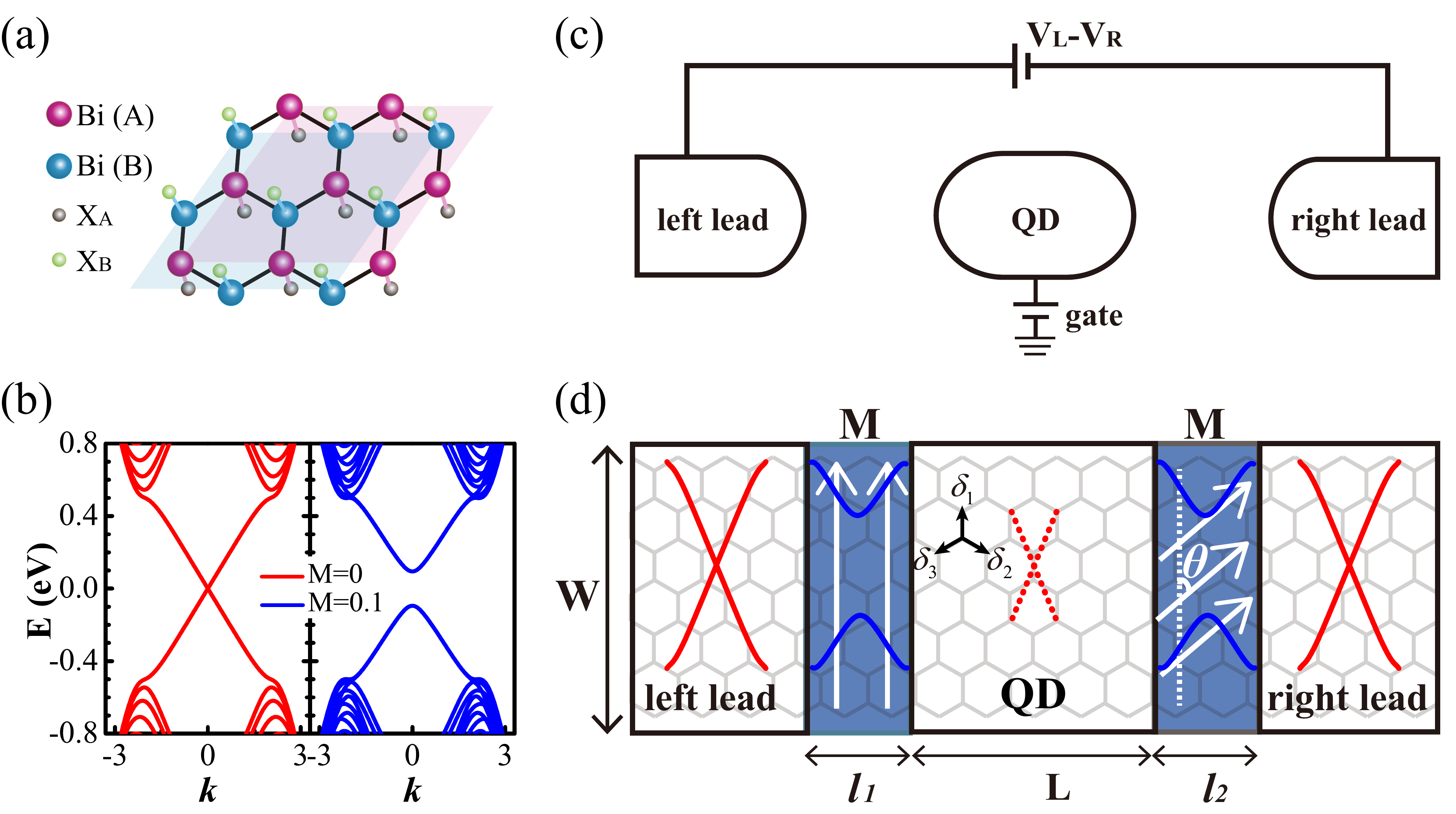}
\caption{\label{figure1} (a) Schematic of bismuthene with a buckled honeycomb structure. The red and blue balls stand for  ${\rm Bi(A)}$ and ${\rm Bi(B)}$ atoms.  The adatoms ${\rm X_A/X_B}$ are adsorbed on the top/bottom side of the Bi monolayer. For proper adatoms (e.g. ${\rm F}$ or ${\rm N}$), a planar magnetization with an exchange field ${\rm M_A/M_B}$ is induced. The orientations of the exchange field can be controlled by the external magnetic field. (b) Two typical band structures of a bismuthene ribbon, which obtained from the tight-binding model. It is gapless (gapped) without (with) the exchange field. The exchange field is set as ${\rm M_A=M_B=M=0.1}$ eV. (c)  Sketch of a typical QD device. A bias  ${\rm V_L-V_R}$ is applied between the left and the right leads. The Fermi energy of the QD can be tuned by the gate voltage. (d) The schematic diagram of a topological QD in a zigzag edged bismuthene ribbon.  In the two dark regions of the ribbon, the planar magnetization induces an exchange field ${\rm M}$. The region has gapped band structures, acting as a depleted region. The energy bands of the two lateral sides of the ribbon remain gapless and they serve as two leads. In the central region, the topological edge states are confined to a series of discrete energy levels, working as a QD. The ribbon width $W=3$ ($ \approx 2.9$ nm). The lengths of the magnetization areas $l_{1}=l_{2}=2$ ($ \approx 1.1$ nm). And the length of the QD $L=5$ ($ \approx 2.75$ nm). $\delta_1$, $\delta_2$ and $\delta_3$ are the three nearest-neighbor vectors.  }
\end{figure*}

Quantum dot (QD) is one of the most important devices in mesoscopic  physics\cite{RHansonLP, SMReimannM, PFendleyAW,PAMaksymT, GBurkardDLoss, JRPetta, YAlhassid}. Because of the promising application for quantum computation and quantum information, engineering QDs with topological states draws lots of interest\cite{AImamogluDD, DLossDP, GBurkard, KChang, PBonderson}. A typical QD system is shown in Fig. \ref{figure1}(c), where the QD is separated from two leads by insulated barriers. The QD has discrete energy levels due to the finite-size effect, and thus carriers can pass through the QD by resonance tunneling. The techniques for fabricating traditional topological QDs depend on chemical etching\cite{GTian,GFulop} or gate voltage depletion\cite{JRPetta,YPSong}. However, such techniques require precise micromachining process and the control is much difficult. To this end, novel approaches for QD engineering are expected. Due to the spin-momentum locked nature of helical edge states in the QSH phase, the planar magnetization opens a gap as large as the exchange field in the energy band of edge states\cite{MKonigHBuhmann, JLLadoJ,DLWangZY,XLQiTL}. And recently, the QSH effect under local planar magnetization has been studied in monolayer systems\cite{ZLiuGZhao,XTAn}. It is found that the tunneling coupling between the separated sides is exponentially weakened by increasing the length of the magnetization area\cite{XTAn}. A natural question is whether we can take such an advantage of planar magnetization in group-V monolayers, which have the large topological bulk gap and highly tunable physical properties, to engineer topological QDs and manipulate their transport properties.

In this paper, we propose a QD in bismuthene nanoribbon by introducing two depletion regions with planar magnetization [as shown in Fig. \ref{figure1}(d)]. The coupling between the QD and the leads on both sides can be controlled by the magnetization. By varying the Fermi level, a series of peaks emerge in zero-temperature differential conductance, which comes from the discrete energy levels of the QD. We find the discrete energy levels originate from the quantum confinement of topological edge states and can still be observed even in a QD with a quite large size. Significantly, the angle between two planar magnetization orientations tunes the positions of discrete energy levels in the topological QD effectively. Through both numerical simulations and  theoretical analysis, the unique confinement mechanism of the topological edge states is revealed. The magnetization-controlled discrete energy levels provide another way to manipulate the spin transport properties of the topological QD. Finally, both  characteristics of electrical conductance and spin conductance of the QD device under finite bias are obtained.

The paper is organized as follows. Section \ref{model} introduces the model of the topological QD and the numerical method to calculate the electrical and spin differential conductance. The key characteristics of the topological QD and its related transport properties are given in Sec. \ref{discussion}. Then a brief conclusion is presented in Sec. \ref{conclusion}.

\section{Model and methods}\label{model}
For convenience, we use zigzag bismuthene ribbons as a platform to demonstrate our proposal in detail. The topological QD device as illustrated in Fig. \ref{figure1}(d) is considered. Two planar magnetization areas (dark regions) separate the device into the QD and the contacting leads. The device can be described by the following Hamiltonian\cite{TZhou}:
\begin{eqnarray} \label{H}
H=&&H_0+H_1~,\nonumber\\
H_0=&&\sum_{{\boldsymbol {i}}\in A,B}~ \sum_{{j}=1}^{3} c_{\boldsymbol {i}}^+~T_{{\boldsymbol {\delta}_j}}~c_{{\boldsymbol {i}}+{\boldsymbol {\delta}_j}}+h.c.\nonumber\\
&&+\sum_{\boldsymbol {i}}c_{\boldsymbol {i}}^+[\lambda_{SO}~\tau_z\otimes \sigma_z]c_{\boldsymbol {i}}~~,\nonumber\\
H_1=&&\sum_{\boldsymbol {i}}c_{\boldsymbol {i}}^+[{\rm M_{A/B}}~ \tau_0 \otimes ({\rm cos\theta}~\sigma_x + {\rm sin\theta}~ \sigma_y )]c_{\boldsymbol {i}}~~.
\end{eqnarray}
Here, $c_{\boldsymbol {i}}$ $(c_{\boldsymbol {i}}^+)$ is the annihilation (creation) operators of electrons at site ${\boldsymbol {i}}$. $\tau$ and $\sigma$   are the Pauli matrices acting in orbital [ $| \phi_+\rangle=-\frac{1}{\sqrt{2}}(p_x+ip_y)$ and $| \phi_-\rangle=\frac{1}{\sqrt{2}}(p_x-ip_y)$ ] and spin space ($\uparrow$ and $\downarrow$), respectively. The hopping $ T_{{\boldsymbol {\delta}_j}}=\begin{pmatrix}
t_1 &{z^{(3-j)}t_2} \\
{z^jt_2} &t_1
\end{pmatrix} \otimes \sigma_0$ describes the nearest hopping from site ${\boldsymbol {i}}$ to ${\boldsymbol {i}}+{\boldsymbol {\delta}_j}$, where $z=e^{i\frac{2\pi}{3}}$ is a constant and $t_{1/2}$ is the hopping coefficient. $\lambda_{SO}$ is the intrinsic spin-orbit coupling strength. ${\rm M_A/M_B}$ refers to the planar exchange field in the ${\rm A/B}$ sublattice, which originates from the adatoms ${\rm X_A/X_B}$. ${\rm M_A/M_B}$ only acts on the planar magnetization regions and equals to zero in other regions. $\theta$ defines the orientation of planar magnetization, which can be controlled by the external magnetic field. Although it has not been observed in experiment yet, the proper adsorption (e.g., ${\rm N}$ or ${\rm F}$ ) can indeed induce large planar magnetization through the first-principle calculations.  Detailed results are provided in the Appendix\cite{footnote1,footnote2}.

To characterize the properties of the topological QD, the zero-temperature electrical differential conductance $G(E)$ at the Fermi energy $E$ is calculated. Based on nonequilibrium Green's function method and Landauer-B${\rm\ddot{u}}$ttiker formula\cite{SDatta}, $G(E)$ under a small bias can be expressed as:
\begin{equation}\label{spinH}
G(E)=\frac{e^2}{h}Tr[\Gamma_L(E){\bf G}^r(E)\Gamma_R(E){\bf G}^a(E)],
\end{equation}
where $\Gamma_p(E)=i[\Sigma^r_p(E)-\Sigma^a_p(E)]$ is the linewidth function of the leads ($p=L,R$) and $ {\bf G}^r(E)=[{\bf G}^a(E)]^\dagger=1/[(E-H_{cen})-\Sigma^r_L-\Sigma^r_R]$ is the retarded Green's function \cite{SDatta}.  $H_{cen}$ contains the Hamiltonian of the QD and two planar magnetization areas. The self-energy $ \Sigma_p^r$ of the semi-infinite lead-$p$ can be calculated numerically\cite{DHLee,MPLopezSancho1,MPLopezSancho2}.

Since spin is a good quantum number in both leads, one can also calculate the spin differential conductances $G_{\uparrow}$ and $G_{\downarrow}$. In analogy to Eq.(\ref{spinH}), $G_{\alpha}(E)$ is obtained by
\begin{equation}
G_{\alpha}(E)=\frac{e^2}{h}Tr[\Gamma_{L\alpha}(E){\bf G}^r(E)\Gamma_{R}(E){\bf G}^a(E)],
\end{equation}
where $\alpha=\uparrow,\downarrow$. Here, $\Gamma_{L\alpha}(E)$ are the corresponding spin part of $\Gamma_{L}(E)$.

\section{Results and Discussions}\label{discussion}

Before presenting our main results, the parameters adopted in the following studies are given. Here, $t_1=1$ eV, $t_2=-1$ eV and $ \lambda_{SO}=0.5$ eV, which are comparable to the fitting parameters of bismuthene from the first-principles data. The magnitude ${\rm M_A/M_B}$  and the orientation angle $\theta$ of the exchange fields can be engineered by the atomic adsorption (e.g., concrete adatoms and their concentration) and the external magnetic field, respectively. For simplicity, we assume the atomic adsorptions are the same in two magnetization areas. Thus, the magnitude of exchange fields (${\rm M_A}$ and ${\rm M_B}$) are the same in two areas. Then, topological helical edge states exist in the low energy region $ E\in[-0.5,0.5]$. The exchange fields in the planar magnetization areas open a gap in the edge states [shown in the right panel of Fig. \ref{figure1}(b)], while its orientation does not change the band structure.

\subsection{A versatile platform for building topological QD devices with tunable key parameters}\label{secA}

We demonstrate our proposal through the simulations of transport properties. Transport performance is not only one of the most essential properties of QD devices, but also an effective method to characterize them. For example, for a non-interacting QD device, the tunneling coupling strength and the discrete energy levels are the key parameters. These parameters can be extracted from the zero-temperature two-terminal differential conductance:
\begin{equation} \label{JaohoH}
G(E)=\frac{ 2e^2}{h} \frac{{\mathsf \Gamma}_L {\mathsf\Gamma}_R}{(E-\varepsilon_d)^2+\frac{1}{4}({\mathsf \Gamma}_L+{\mathsf \Gamma}_R)^2 },
\end{equation}
where $E$ denotes the Fermi energy, $\varepsilon_d$ denotes the discrete energy, ${\mathsf \Gamma}_L$ $({\mathsf \Gamma}_R)$ labels the tunneling coupling strength between the QD and the left (right) leads \cite{HHaugAPJauho}.

\begin{figure}
\includegraphics[width=\columnwidth]{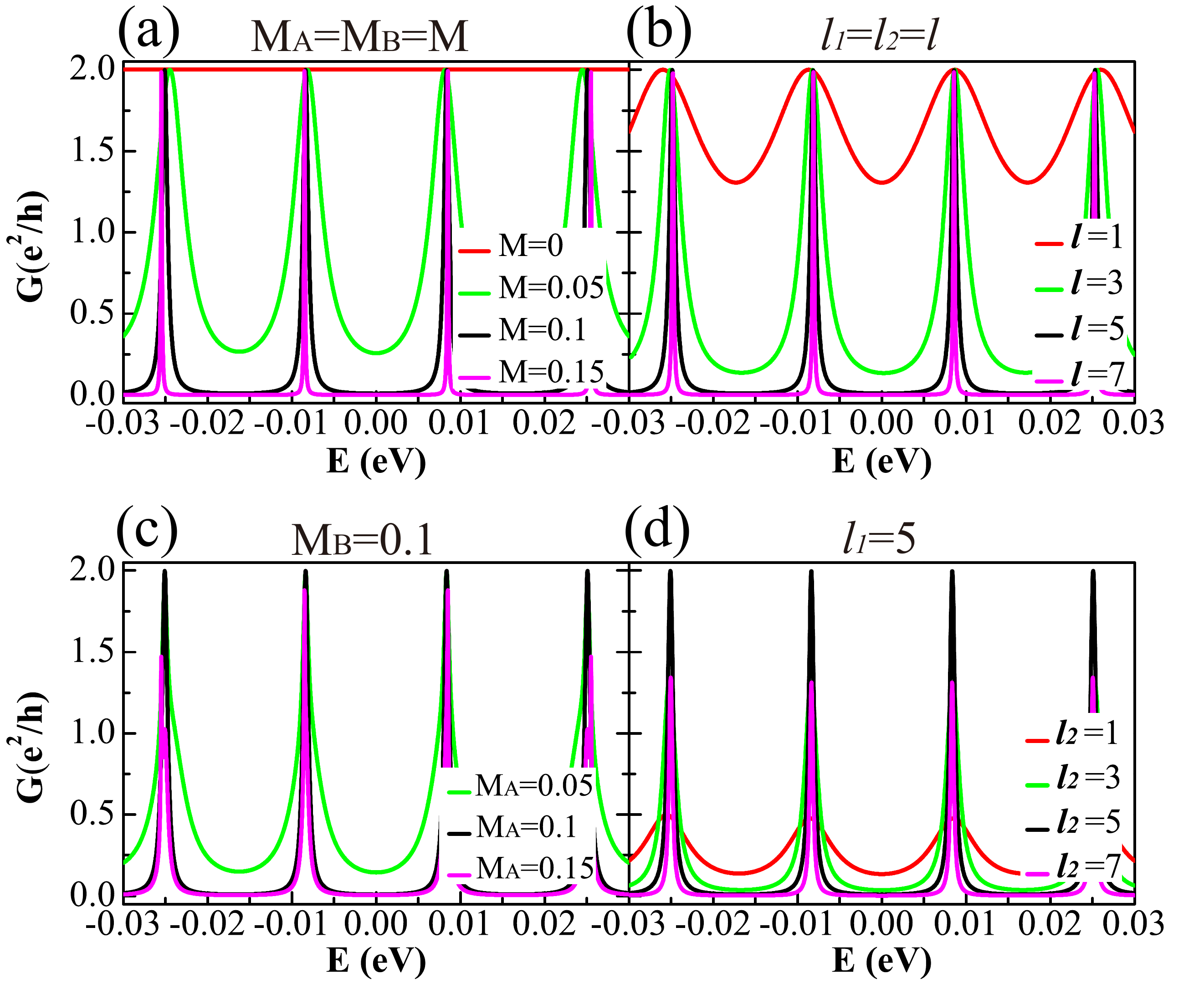}
\caption{\label{figure2} $G$-$E$ relations of the QD device for  different exchange fields of A/B sublattice ${\rm M_A/M_B}$ [(a) and (c)] and length of the magnetization areas $l_{1}/l_{2}$ [(b) and (d)]. In (a) and (c),  $l_{1}=l_{2}=l=5$, In (b) and (d),  ${\rm M_A=M_B=M}=0.1$ eV.  Other parameters are $W=10$, $L=50$ and $\theta=0$.}
\end{figure}

In Fig. \ref{figure2}(a) and \ref{figure2}(b), the linear differential conductance $G$ versus the Fermi energy $E$ under different exchange fields ${\rm M_A=M_B=M}$ and lengths of the magnetization area $l_{1}=l_{2}=l$ are plotted. In the absence of planar magnetization (${\rm M}=0$), $G\equiv 2e^2/h $ indicates no QD is found in the device [see the red line in Fig. \ref{figure2}(a)]. However, when ${\rm M} \neq 0$ and $l \neq 0$, $G$ oscillates from $0$ to $2e^2/h $ in the energy region $E \in [-0.03, 0.03]$, indicating the formation of the QD by introducing the two magnetization areas. The energies of peaks stand for the positions of the discrete energy levels. The sharpness of peaks represents the tunneling coupling strength between the QD and the leads. The physical picture for the QD is very intuitive. The planar magnetization can open a edge state gap by adding two barriers that locate on corresponding magnetization regions [see dark regions in Fig. \ref{figure1}(d)]. The electronic states inside the central region are confined to the discrete energy levels of the QD sample. Due to the finite width and height of the two barriers, the carriers can pass through the device by quantum tunneling. If the Fermi energy $E$ equals the discrete energy level $\varepsilon_d$, the resonance tunneling happens. Therefore, a proposal for building a QD device from a bismuthene nanoribbon is established.

The proposed QD device is a versatile platform. Its key parameters can be easily tuned. First, we show the feasibility to control the tunneling coupling strength and the related transport properties. Physically, the tunneling coupling strengths between the QD and two leads are determined by the probability of quantum tunneling across the planar magnetization regions. It can be changed by the exchange fields (${\rm M_A}$ and ${\rm M_B}$) and the lengths of the magnetization area ($l_{1}$ and $l_{2}$). In Fig. \ref{figure2}(a) and \ref{figure2}(b), one can find that the oscillations are weak for small ${\rm M}$ and $l$ [e.g., ${\rm M}=0.05$ eV in Fig. \ref{figure2}(a), $l=1$ in Fig. \ref{figure2}(b)]. By increasing ${\rm M}$ or $l$, the tunneling coupling strength decays exponentially. $G$ shows clear oscillations from $0$ to $2e^2/h$ and their peaks become narrow. For large ${\rm M}$ and $l$, the quantum tunneling from the leads to the QD is difficult. Correspondingly, the peaks of $G$ are too sharp to be observed [e.g., ${\rm M}=0.15$ eV in Fig. \ref{figure2}(a), $l=7$ in Fig. \ref{figure2}(b)]. Besides, $G$ versus $E$ for ${\rm M_A \neq M_B}$ and $l_1 \neq l_2$ are also studied. In Fig. \ref{figure2}(c), by varying ${\rm M_A}$, the sharpness of $G$ is changed, while the maximum of its peaks nearly remains $2e^2/h$. In contrast, when $l_2=1$, not only the sharpness of $G$ is changed, but also its peak maximums decreases from  $2 e^2/h$ to $0.5 e^2/h$ [see Fig. \ref{figure2}(d)]. The reason is intuitively simple. As stated in Eq. (\ref{JaohoH}), when $l_1=l_2$, the tunneling coupling strength of two peaks are equal, i.e., ${\mathsf \Gamma}_L ={\mathsf \Gamma}_R$. A peak of $2e^2/h$ is obtained.  While $l_1 \neq l_2$, one obtains a smaller resonant peak due to ${\mathsf \Gamma}_L \neq {\mathsf \Gamma}_R$. For example, when $l_2=1$ and $\Gamma_R$ is nearly 14 times as $\Gamma_L$, the resonant conductance gives $G\approx 0.5e^2/h$.

\begin{figure}
\includegraphics[width=9 cm]{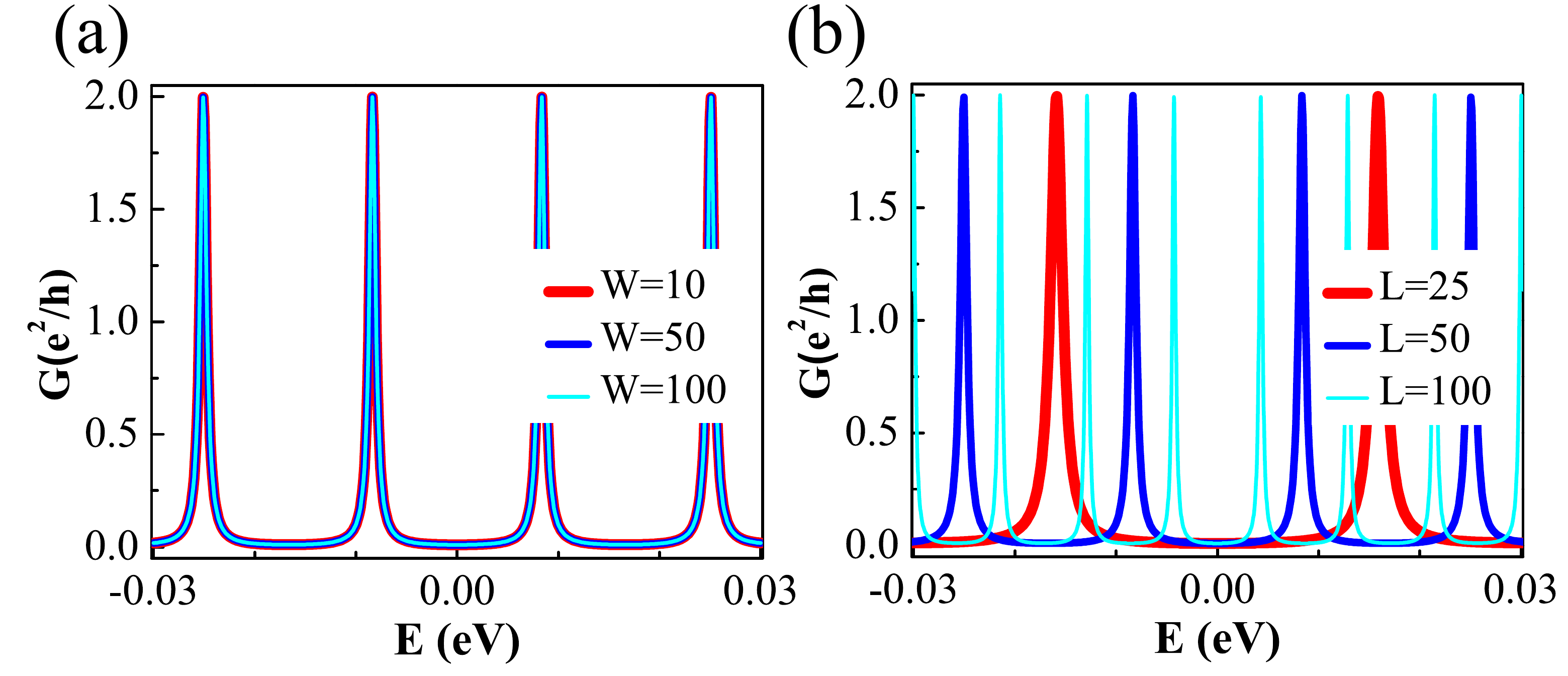}
\caption{\label{figure3}   $G$ versus  $E$ for different  width $W$ (a) and length $L$ (b) of the bismuthene QD. The parameters are set as ${\rm M_A=M_B}=0.1$ eV, $l_{1}=l_{2}=5$ and $\theta=0$. In (a), $L=50$, and in (b),  $W=10$.}
\end{figure}

The discrete energy level is another key parameter of the QDs.  In traditional QDs, the discrete energy levels are sensitive to its size. An increasing size always generates more discrete energy levels. In our proposal, the QD is built from a topological material. What are the differences between the traditional and our topological QD? In the following, we study it in detail. For better observation of the discrete energy levels, we use a symmetric model, ie., ${\rm M_A=M_B}$, $l_{1}=l_{2}$ and $\theta=0$.

The differential conductance $G$ versus the Fermi energy $E$ for different ribbon widths $W$ and lengths $L$ are plotted Fig. \ref{figure3}. One can find that the conductance $G$ is not affected by $W$ as the curves of $W=10$, $50$ and $100$ coincide with each other. It manifests that the discrete energy levels come from the quantum confinement of the topological edge states and the number of the discrete energy levels cannot be changed by the width of the QD. 
In contrast, the length $L$ can effectively manipulate the peaks of the conductance, i.e., discrete energy levels of the QD. As shown in Fig. \ref{figure3}(b), the increasing length $L$ induces more discrete energy levels.  Further, the spacing of discrete energy levels is equal and inversely proportional to the length $L$. This phenomenon is also consistent with the linear dispersion of topological edge states. The behaviors of $G$ versus both $L$  and $W$ not only prove that the discrete energy levels stem from the confinement of the topological edge states, but also demonstrate that the discrete energy levels can also be controlled.

It is worth noting that, different from the traditional QDs where discrete energy levels come from the confinement of bulk states, the absence of bulk states makes the spacing of discrete energy levels much larger in the present topological QD. For example, one can estimate from Fig. \ref{figure3}(b) that the level spacing is about 10 meV for the topological QD with size $55~{\rm nm}$ ($L \approx 100 $). The immunity of discrete energy levels to the width makes the fabrication and the observation of the topological QDs in experiments more easily.

\subsection{Manipulate the discrete energy levels by the angle of planar magnetization orientations}
In the subsection \ref{secA}, we have proposed a topological QD device in the bismuthene nanoribbon. We demonstrate that its key parameters can be tuned by the barriers and the sample size. These characteristics also exist more or less in the traditional or other topological QD devices. Does any unique topological property exist in our topological QD? In this subsection, we show the discrete energy levels can be adjusted by the angle $\theta$ between two planar magnetization orientations.

\begin{figure}
\includegraphics[width=\columnwidth]{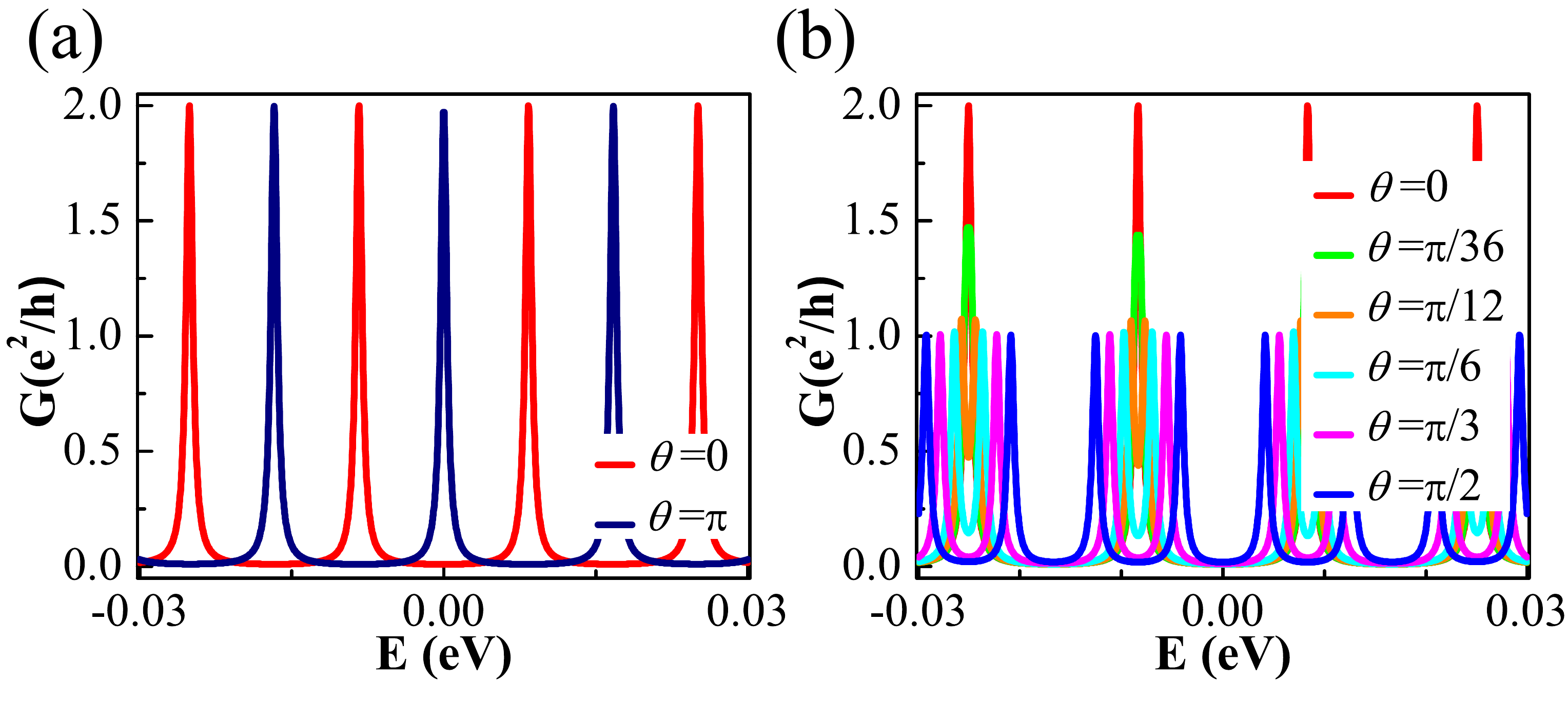}
\caption{\label{figure4} $G$-$E$ relations for different angles $\theta$ between two planar magnetization orientations. (a) $\theta=0$ and $\theta=\pi$. (b) $\theta$ changes from $0$ to $\pi/2$ gradually. Other parameters are ${\rm M_A=M_B}=0.1$ eV, $l_{1}=l_{2}=5$, $ L=50$ and $W=10$.}
\end{figure}

In Fig. \ref{figure4}, we study the behaviors of the differential conductance $G$ under different $\theta$. Although the orientation of the exchange fields cannot change the insulating nature of the potential barriers, surprisingly, it can regulate the discrete energy levels of the QD. When two planar magnetization orientations change from being parallel ($\theta=0$) to being antiparallel ($\theta=\pi$), the conductance peaks for $\theta=\pi$ appear in the middle of the neighbor conductance peaks for $\theta=0$ [see Fig. \ref{figure4}(a)]. In other words, the discrete energy levels shift a half period by reversing a planar magnetization orientation. This property may be used as a quantum bit with two states $0$ and $1$, in which the transition condition is just reversing the magnetic field in one planar magnetization area. When the angle $\theta$ doesn't equal to $0$ or $\pi$, the linear conductance cannot reach $2e^2/h$. The peaks split and the amplitudes decrease from $2e^2/h$ to $e^2/h$ when the angle $\theta$ increases from $0$ to $\pi/2$. And the peaks will restore to $2e^2/h$ gradually when $\theta$ changes from $\pi/2$ to $\pi$.   The former variation process is given in Fig. \ref{figure4}(b). In order to explain this phenomenon, we speculate that the factor $2$ originates from two edge states. The discrete energy levels, arising from the confinement of topological helical states at the upper or lower edge of the QD, contribute a conductance peak with an amplitude $e^2/h$. And the angle $\theta$ shifts discrete energy levels to the opposite energy directions.  In the following, two evidences are provided to support the argument.

\begin{figure}
\includegraphics[width=\columnwidth]{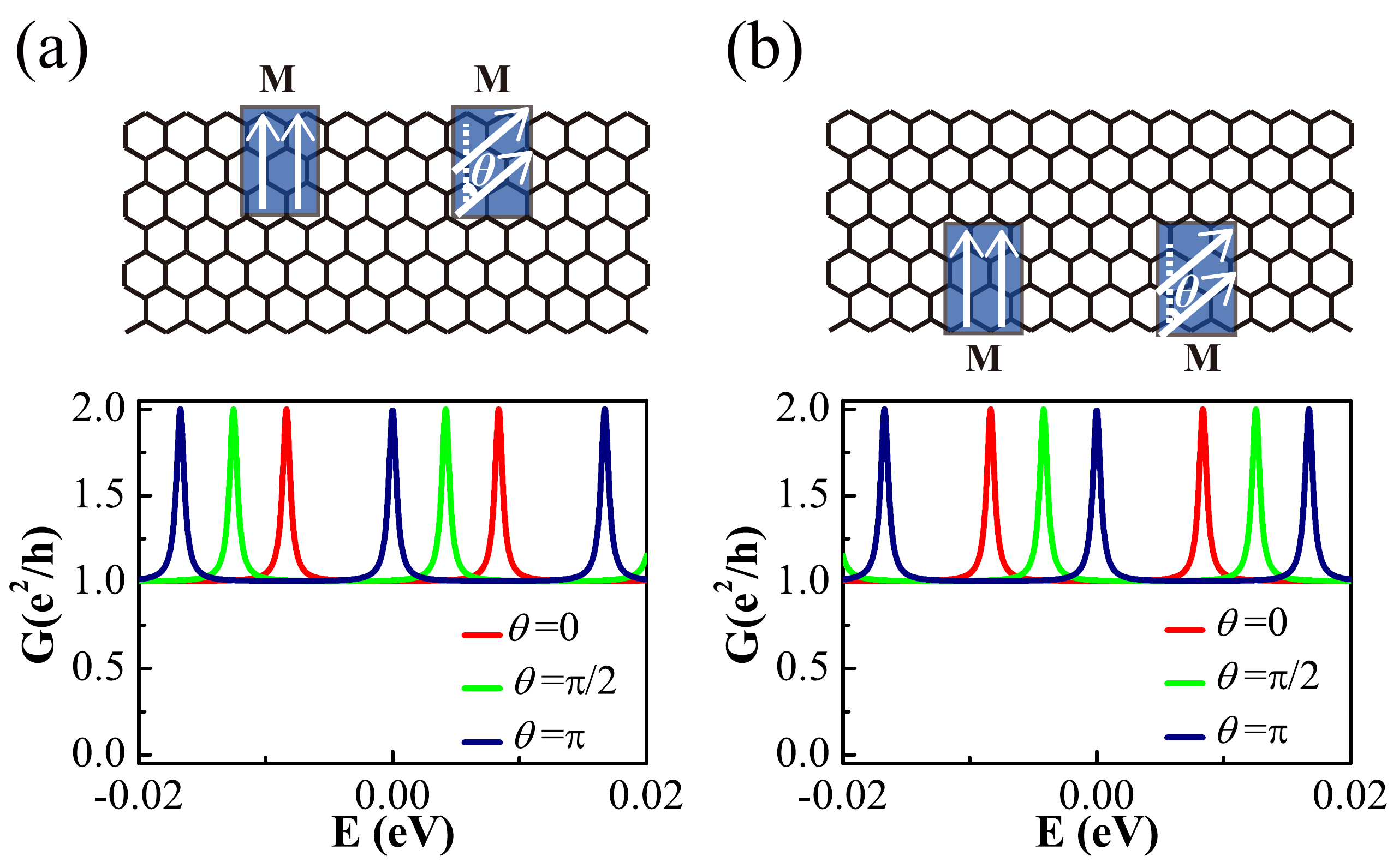}
\caption{\label{figure5} $G$-$E$ relations under different angles $\theta$ when the exchange field ${\rm M}$ only exists at the upper edge (a) or the lower edge (b) of the bismuthene ribbon. The parameters are the same as those in Fig. \ref{figure4}.}
\end{figure}

The first evidence is given by the conductance simulation of two modified setups. As shown in Fig. \ref{figure5}, the exchange field is only introduced at one edge.  When the exchange fields exist at the upper edge, the helical edge state at the lower edge are dissipationless and contributes to a quantized conductance $e^2/h$. The confined states at the upper edge make the differential conductance $G$ oscillate between $e^2/h$ and $2e^2/h$. And the peaks of $G$  shift periodically with the angle $\theta$. We plot the results of $G$ under $\theta=0,~\pi/2,~\pi$, and find that the evenly spaced peaks shift leftwards [see the bottom panel of Fig. \ref{figure5}(a)]. When the exchange fields exist at the lower edge, the peaks of $G$ also oscillate from $e^2/h$ to $2e^2/h$. But the evenly spaced peaks of $G$ under $\theta=0,~\pi/2,~\pi$ shift rightwards. For $\theta=0$ and $\theta=\pi$, the conductance peaks contributed by the upper edge and lower edge meet at the same energy positions. So $G$ has a oscillation from $0$ to $2e^2/h$ if the exchange fields exist at the both edges. When $\theta$ deviates from those two specific angles, the degeneracy of discrete energy levels at the two edges are broken and each peak of $G$ splits into two. Thus, the amplitude of $G$ decreases from $2e^2/h$ to $e^2/h$ [see Fig. \ref{figure4}(b)].

The second evidence is given by the theoretical analysis of the bound energies, i.e., discrete energy levels,
at two edges. The proposed QD structure can be simplified as a one-dimensional finite potential well with a length $L$ and two semi-infinite barriers with the mass term ${\rm M}$. The upper edge states inside the potential well can be described by the Hamiltonian
\begin{equation}\label{indot}
\mathcal{H_{QD}}=\begin{pmatrix}
\hbar vk & 0\\
0& -\hbar vk
\end{pmatrix}.\end{equation}
Here, the operator $k=-i \frac{\partial}{ \partial x}$ and $v$ denotes the Fermi velocity. The eigenfunction in the range $x\in[0,L]$ can be expressed as
\begin{equation}\label{indotf}
\Phi_\mathcal{{QD}}(x)=\frac{{\mathcal A}}{\sqrt{v}}\binom 1 0 e^{i\frac{E}{\hbar v}x}+\frac{{\mathcal B}}{\sqrt{v}}\binom 0 1 e^{-i\frac{E}{\hbar v}x},
\end{equation}
with the bound energy $E$.

In the left side, the barrier is ${\rm M} \sigma_x$ and the Hamiltonian $\mathcal{H_L}$ can be written as
\begin{equation}\label{left}
\mathcal{H_L}=\begin{pmatrix}
\hbar vk & {\rm M}\\
{\rm M}& -\hbar vk
\end{pmatrix}.\end{equation}
Here, an evanescent wave with vector ${\rm k_1}=-\frac{i\sqrt{{\rm M}^2-E^2}}{\hbar v}$ ($E<{\rm M}$) exists in the barrier. The corresponding eigenfunction in the range $x\in (-\infty,0)$ is
\begin{equation}\label{leftf}
\Phi_\mathcal{L}(x)=\frac{{\mathcal C}}{\sqrt{v}}\binom {{\rm M}} {E+i\sqrt{{\rm M}^2-E^2}} e^{\frac{\sqrt{{\rm M}^2-E^2}}{\hbar v} \ x}.
\end{equation}

\begin{figure}
\includegraphics[width=\columnwidth]{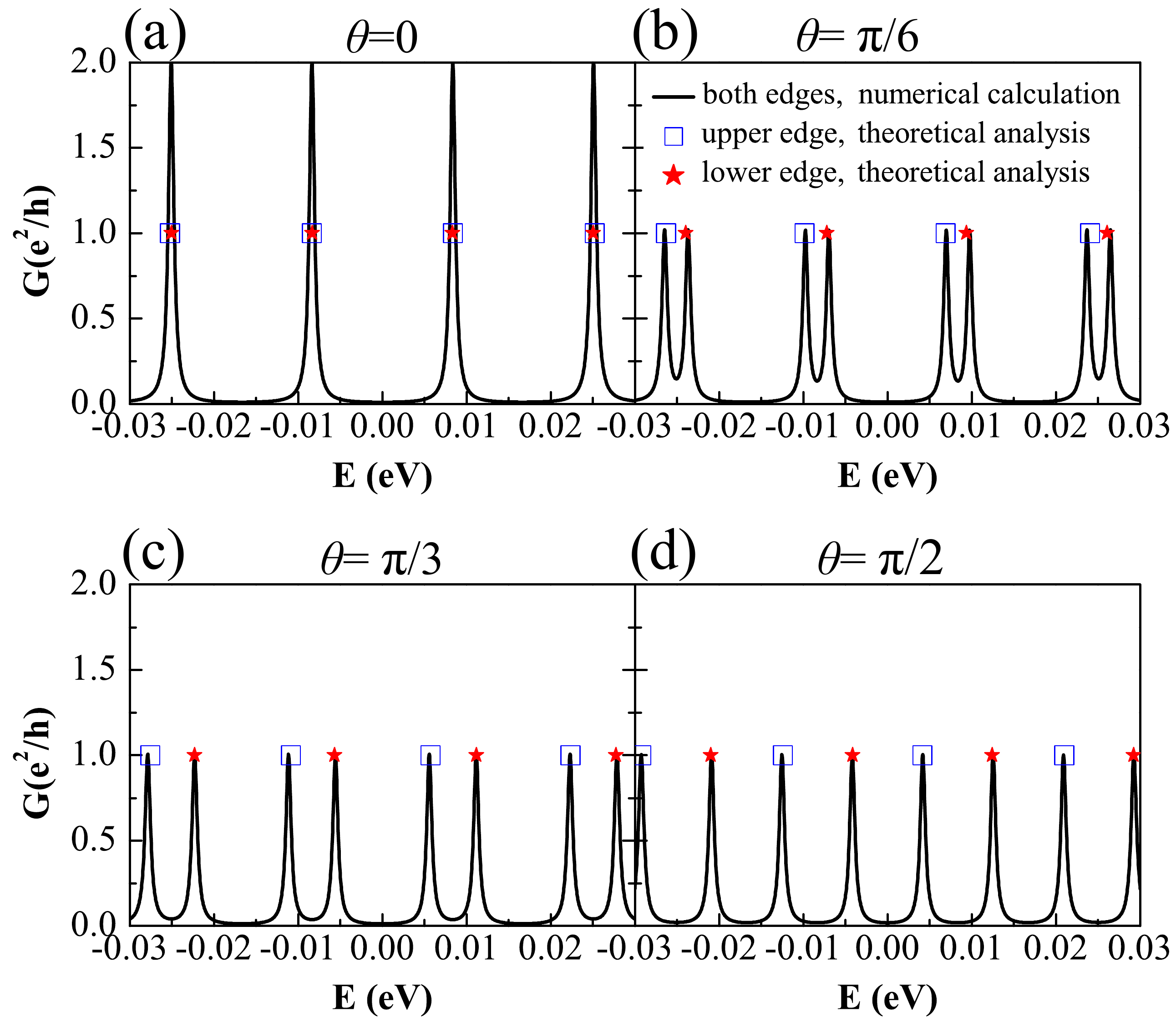}
\caption{\label{figure6} A comparison of discrete energy levels from the numerical simulation and the theoretical analysis. The numerical results  are extracted from Fig. \ref{figure4}(b), and the analytical results are the solutions of the transcendental Eqs. (\ref{E1}) and (\ref{E2}).}
\end{figure}

In the right side, the barrier is ${\rm M }\cos\theta~\sigma_x+{\rm M}\sin\theta~\sigma_y$ and the Hamiltonian $\mathcal{H_R}$ is
\begin{equation}\label{right}
\mathcal{H_R}=\begin{pmatrix}
\hbar vk & {\rm M}e^{-i\theta}\\
{\rm M}e^{i\theta}& -\hbar vk
\end{pmatrix}.
\end{equation}
Here, the rightward evanescent wave vector is ${\rm k_2}=\frac{i\sqrt{{\rm M}^2-E^2}}{\hbar v}$ and the corresponding eigenfunction in the range $x\in (L,\infty)$ is
\begin{equation}\label{rightf}
\Phi_\mathcal{R}(x)=\frac{{\mathcal D}}{\sqrt{v}}\binom {{\rm M}e^{-i\theta}} {E-i\sqrt{{\rm M}^2-E^2}} e^{-\frac{\sqrt{{\rm M}^2-E^2}}{\hbar v} \ x}.
\end{equation}
According to the continuity condition $\Phi_\mathcal{QD}(0)=\Phi_\mathcal{L}(0)$, $\Phi_\mathcal{QD}( L)=\Phi_\mathcal{R}(L)$, we obtain the relationship for the bound energy  at the upper edge, that is,
\begin{equation}\label{E1}
e^{i(2\frac{E}{\hbar v}L+ \theta)}= \frac{E+i\sqrt{{\rm M}^2-E^2}}{E-i\sqrt{{\rm M}^2-E^2}}.
\end{equation}

The lower edge states inside the potential well can be described by Hamiltonian
\begin{equation}\label{indot}
\mathcal{H_{QD}}=-\begin{pmatrix}
\hbar v k & 0\\
0& -\hbar v k
\end{pmatrix}.\end{equation}
The potential in the two sides are still ${\rm M} \sigma_x$ and ${\rm M}\cos\theta~\sigma_x+{\rm M}\sin\theta~\sigma_y$. After some  algebra, one can obtain  the  corresponding relationship for  the bound energy at the lower edge:
\begin{equation}\label{E2}
e^{i(2\frac{E}{\hbar v}L-\theta)}= \frac{E+i\sqrt{{\rm M}^2-E^2}}{E-i\sqrt{{\rm M}^2-E^2}}.
\end{equation}

A comparison of the numerical simulation and the theoretical analysis is given in Fig. \ref{figure6}. The numerical results are extracted from Fig. \ref{figure4}(b), and the analytical results are the solutions to the Eqs. (\ref{E1}) and (\ref{E2}). The discrete energy levels from the upper edge or the lower edge contribute a quantized conductance $e^2/h$. So we label the solutions of the Eqs. (\ref{E1}) and (\ref{E2})  in the values of $G = e^2/h$. In Fig. \ref{figure6}, all the labels of the solutions locate at the peaks of $G$. Further, when two kinds of labels, i.e., bound energies at the two edges, coincide with each other [see Fig. \ref{figure6}(a)], peaks of $G=2e^2/h$ emerge. If these two kinds of labels separate with each other [see Fig. \ref{figure6}(b-d)], the  peaks of $G$ reduce from $2e^2/h$ to $e^2/h$ and every solution is accompanied with a peak [see Fig. \ref{figure6}(b-d)]. Significantly, comparing four subplots with the increasing angle $\theta$, one finds that the discrete energy levels contributed by the upper edge will move in a negative energy direction, while the discrete energy levels contributed by the lower edge will move in the positive energy direction.

The numerical simulation of differential conductance, the theoretical analysis of bound energies and the prefect coincidence of their results all confirm that the discrete energy levels can be manipulated by the angle between two planar magnetization orientations. From the theoretical analysis, such a manipulation requires: (1) the band structure with helical edge states; (2) the opposite propagation directions for spin carrier between the upper and the lower edges; (3) the special boundary condition induced by the exchange fields. All three characteristics are the essential properties of the QSH phase. Therefore, the manipulation by the angle $\theta$ is unique in the present proposed topological QD.

\subsection{Application of unique manipulation mechanism by the angle of planar magnetization orientations}
We have shown that the angle of the planar magnetization orientations introduces a special confinement mechanism to the topological edge states, leading to the unique manipulation of the discrete energy levels in the proposed QD. It is natural to ask whether we can utilize such mechanism for spintronics. As the bismuthene is a QSH material, the spin-up carriers will flow along the upper edge while the spin-down carriers flow along the lower edge under a small bias. Since the discrete energy levels at upper and lower edges move to different directions by tuning $\theta$. Thus, such a manipulation may have applications in spintronics. We test the application by studying the spin differential conductance $G_{\uparrow}$, $G_{\downarrow}$ and $G_{\uparrow}-G_{\downarrow}$.

\begin{figure}
\includegraphics[width=\columnwidth]{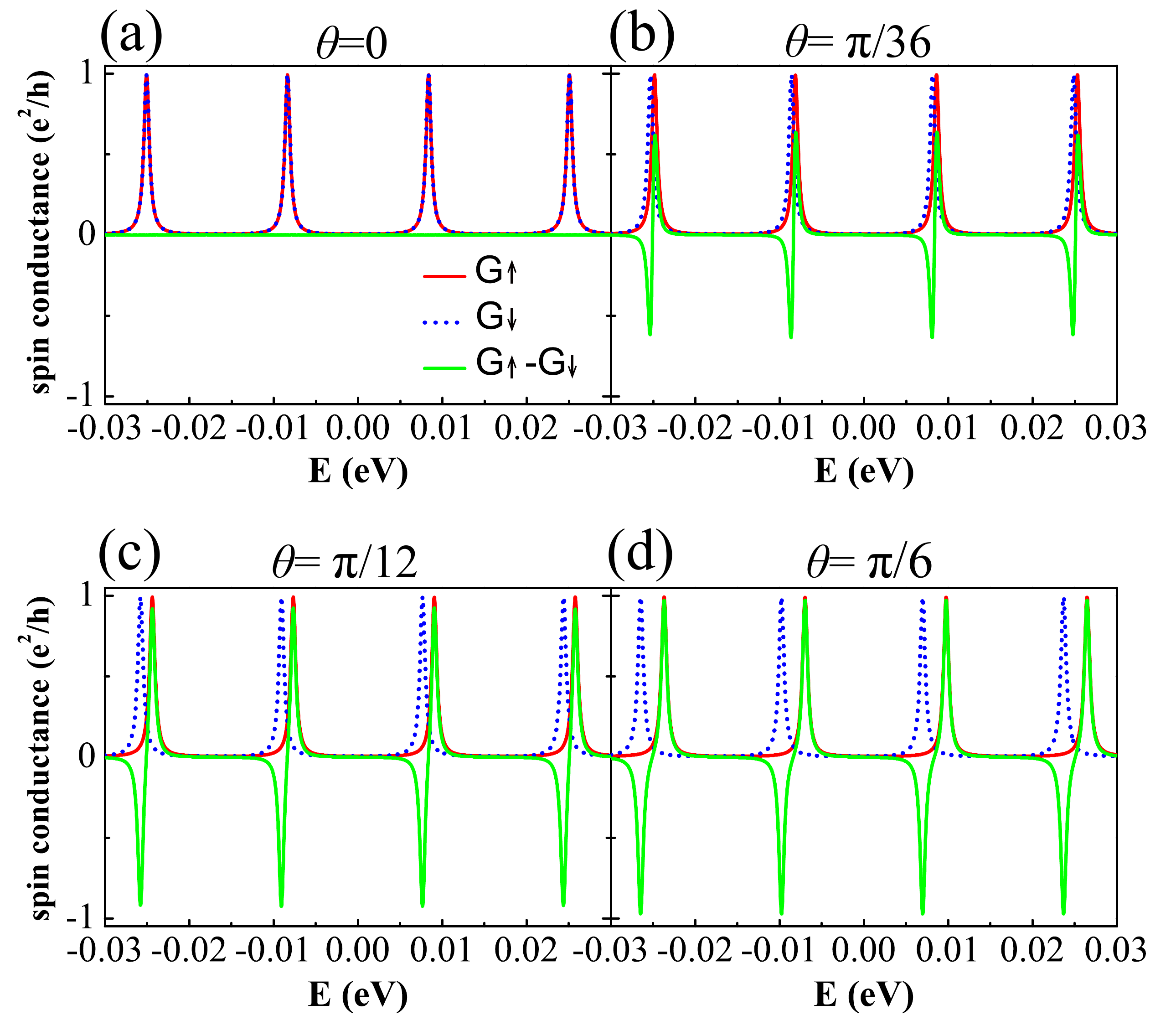}
\caption{\label{figure7} The spin conductance $G_{\uparrow}$, $G_{\downarrow}$ and $G_{\uparrow}-G_{\downarrow}$ versus the Fermi energy $E$  for different angles $\theta=0$ (a), $\theta=\pi/36$ (b), $\theta=\pi/12$ (c) and $\theta=\pi/6$ (d). Other parameters are the  same as those in  Fig. \ref{figure4}.}
\end{figure}

Figure \ref{figure7} plots the variation of spin differential conductance $G_{\uparrow}$, $G_{\downarrow}$ and $G_{\uparrow}-G_{\downarrow}$ with the Fermi energy $E$. Both  $G_{\uparrow}$ and $G_{\downarrow}$ oscillate from $0$ to $e^2/h$ in the proposed QD device. The peaks of $G_{\uparrow}$ and $G_{\downarrow}$ correspond to
the discrete energy levels at the upper and lower edge, respectively. When two planar magnetization orientations are parallel ($\theta=0$), $G_{\uparrow}$ and $G_{\downarrow}$ are identical and their variation $G_{\uparrow}-G_{\downarrow}$ is zero [see the blue line in Fig. \ref{figure7}(a)]. There is no spin current in the QD device. In contrast, by increasing $\theta$, the peaks of $G_{\uparrow}$ and $G_{\downarrow}$ shift to the opposite energy directions, and consequently $G_{\uparrow}-G_{\downarrow}$ becomes nonzero. The current in the QD device becomes spin polarized. For a small $\theta$ [see Fig. \ref{figure7}(b) and \ref{figure7}(c)], the separation of the discrete energy levels coming from upper and lower edges is small, the peaks of  $G_{\uparrow}-G_{\downarrow}$ are smaller than $e^2/h$. This means both the upper and lower edges contribute a finite current, and the current is partially spin polarized. For a large $\theta$ [see Fig. \ref{figure7}(d)], $G_{\uparrow}$ and $G_{\downarrow}$ are well separated in energy. Thus, a pure spin current can be obtained in this case. Moreover, by tuning the Fermi energy $E$, $G_{\uparrow}-G_{\downarrow}$ can be tuned from $-e^2/h$ to $e^2/h$. In other words, the current can be switched from pure spin-up polarization to pure spin-down polarization.

\begin{figure}
\includegraphics[width=\columnwidth]{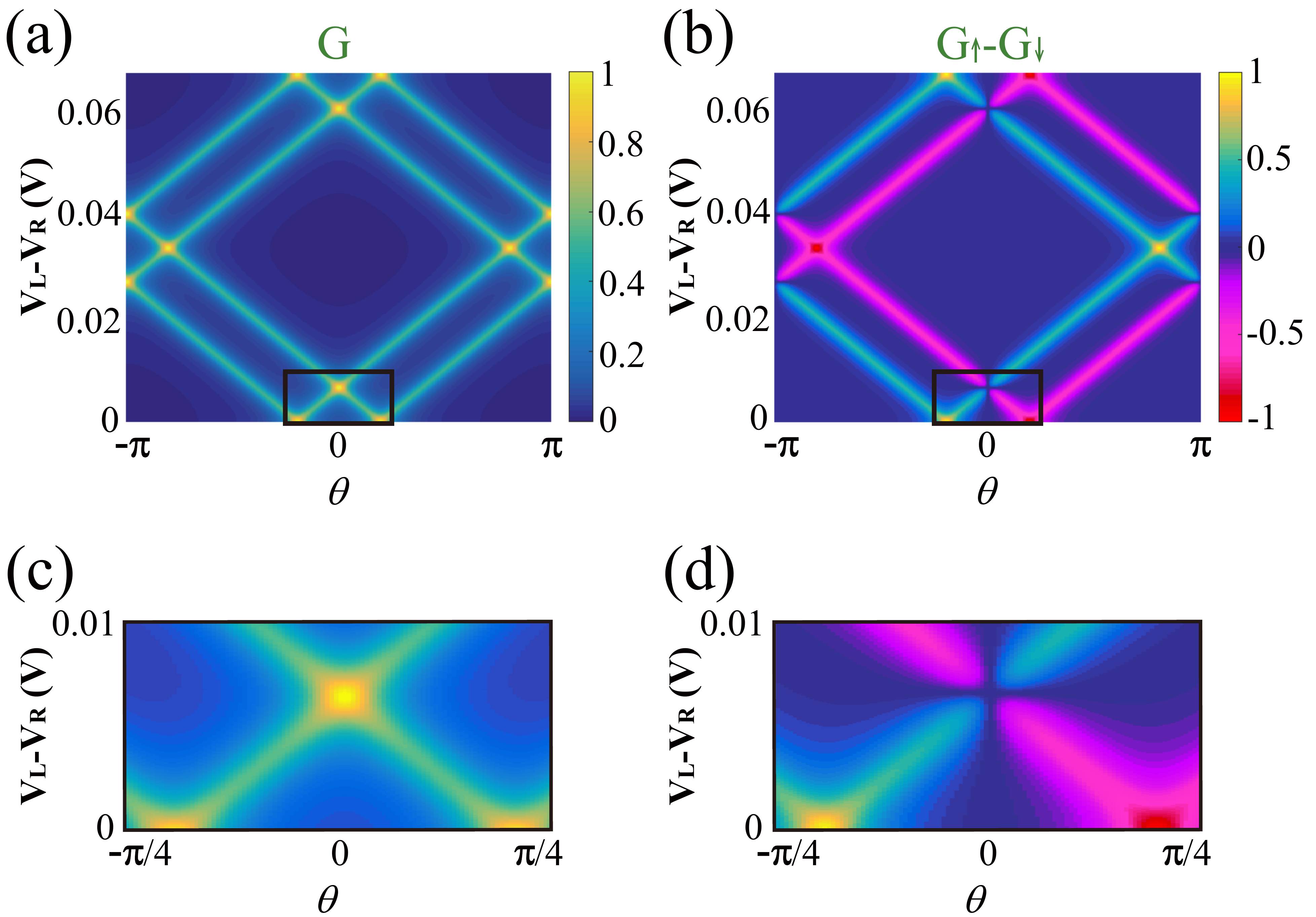}
\caption{\label{figure8} The electrical and spin differential conductance diagrams under the combination of finite bias (${\rm V_L-V_R}$) and the angle $\theta$. (c) and (d) are the zooming in of the selected black box region in (a) and (b). The Fermi energy is $E=-0.01$ eV. Other parameters are the same as those in Fig. \ref{figure4}.}
\end{figure}

Next, we give another method to manipulate the spin polarized current other than by tuning the Fermi energy, as shown in Fig. \ref{figure8}. In a traditional QD, the transport properties are also tuned by a finite bias $V$. Here, we set $V=V_L-V_R$. In this case, the differential conductance and the spin differential conductance are modified to $G(E,V)=[G(E+eV/2)+G(E-eV/2)]/2$ and
$G_{\alpha}(E,V)=[G_{\alpha}(E+eV/2)+G_{\alpha}(E-eV/2)]/2$, respectively. Figure \ref{figure8} is obtained from these two formulas. The bias $V$ plays the similar role as the Fermi energy $E$. By tuning $V$, the resonant tunneling can also be observed in $G$ [see Fig. \ref{figure8}(a) and \ref{figure8}(c)].  $G_{\uparrow}(E,V)-G_{\downarrow}(E,V)$ shows a rapid oscillation by variation of both angle $\theta$ and finite bias $V$, except $\theta=0$ or $\theta =\pi$ [see Fig. \ref{figure8}(b) and \ref{figure8}(d)]. Therefore, the unique manipulation of spin transport by the angle of planar magnetization orientations always holds.

From above studies, one can conclude that the spin transport properties of the proposed topological QD can be controlled by a new parameter, the angle $\theta$ between two planar magnetization orientations. Therefore, such unique manipulation mechanism has promising applications in spintronics.

\section{Conclusion}\label{conclusion}

In this paper, we select bismuthene as a candidate of our proposal. The most important reason is that bismthene is a QSH material with a large bulk gap and has been realized in experiments \cite{FReis}. More recently, the QSH is also observed in ${\rm Na_3Bi}$ with a bulk gap of $0.4$ eV \cite{JLCollinsA}. Besides, lots of materials, such as stanene\cite{FFZhuWJChen,YXuBYan} and ${\rm MoSe_2}$\cite{XFQianJWLiu,YZhangTRChang}, are predicted to host large bulk gap QSH effects. In principle, our proposed topological QD model can also be applied in these systems. To well observe the QD phenomena in experiments, the spacing of the discrete energy levels in QDs prefers one order smaller than the bulk gap. Because of the limited energy resolution, the distinction of discrete energy levels and the application of the topological QD device in small bulk gap QSH materials may be difficult.

In summary, we find a new method to engineer the topological QD system in bismuthene. The QD effect arises from the quantum confinement of the topological edge states by applying planar magnetizations. The coupling strength, the discrete energy levels and other key parameters of the QD can be controlled feasibly. Interestingly, different from the conventional QD, we find that the angle $\theta$ between two planar magnetization orientations can effectively tune the discrete energy levels of this topological QD. The phenomenon originates from the unique confinement mechanism of the topological edge states under different boundary conditions, caused by the variation of angle $\theta$. Finally, we find the spin transport properties of the topological QD can also be manipulated by such a mechanism.

\section{ACKNOWLEDGMENTS}
We thank Haiwen Liu, Qing-feng Sun, Zhi-min Yu and  and Wen-Long You for helpful discussion. This work was supported by NSFC under Grants No. 11534001, 11822407, 11874298, 11574051 and 11874117, NSF of Jiangsu Province under Grants No. BK2016007.   J.J. Zhou and  T. Zhou contributed equally to this work.

\section{Appendix}\label{Appendix}
\begin{figure}
\includegraphics[width=\columnwidth]{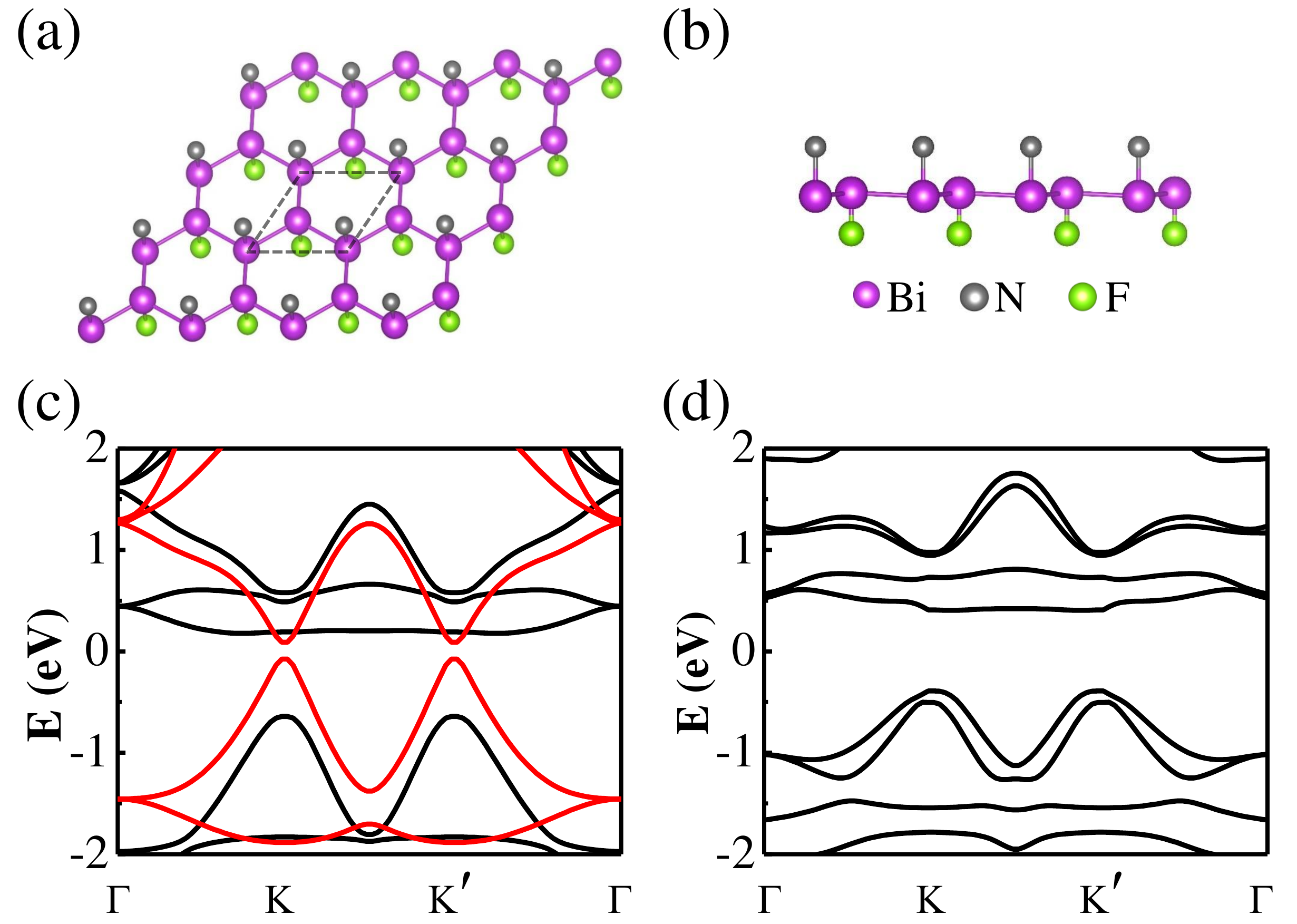}
\caption{\label{figure9} (a) Top and (b) side views of the structure of monolayer Bi$_2$NF. The dashed line in (a) indicates the unit cell of the system. (c) and (d) Energy bands for the monolayer Bi$_2$NF without and with SOC considered, respectively. The red/black curve in (c) indicates the spin-up/spin-down state.}
\end{figure}

To observe the interesting phenomena in our proposal, we need a real material system described by the Hamiltonian Eq. (\ref{H}). Based on first-principles calculations, we found the nitrogenated bismuth fluoride monolayer Bi$_2$NF [shown in Fig. \ref{figure9}(a) and (b)] is a good material candidate for our proposal. The Bi$_2$NF has a similar structure to the bismuth fluoride\cite{ZGSong} but with one side of the F atoms replaced by the N atoms. The N atom induces a net magnetic moment of $2~{\rm\mu_B}$ in the unit cell of Bi$_2$NF, making the system ferromagnetic. Calculated band structure of Bi$_2$NF without spin-orbit coupling shows an obvious splitting of the spin-up and spin-down bands [see Fig. \ref{figure9}(c)]. When SOC is taken into account, magnetic anisotropy energy calculations show that the easy magnetization axis lies in-plane with about $20$ meV lower in energy than the out-plane magnetization. The calculated bands with SOC show that Bi$_2$NF is a ferromagnetic insulator with a gap of $0.8$ eV [see Fig. \ref{figure9}(d)], supporting our proposed Hamiltonian Eq. (\ref{H}). In our model, the comparable fitting  parameters are set as $t_1=1$ eV, $t_2=-1$ eV, $\lambda_{SO} = 0.5$ eV.


\begin{thebibliography}{99}
\bibitem{footnote1}
A staggered potential is induced by the different adatoms in the magnetic areas from the first-principles calculation, but its value is very small and it cannot open a gap in topological edge states. Namely, it take little effect  to the insulated barrier. Thus, it is neglected in our model.
\bibitem{footnote2}
From first-principles calculations, all doped Cr, Mo and W atoms prefer out-plane magnetizations\cite{TZhou1}, but their magnetic anisotropy energies are about $1$ meV per unit cell. As a result, the orientations of the magnetizations can be easily switched to in-plane by the external magnetic field.
\end{thebibliography}
\end{document}